\documentclass[12pt]{article}

\usepackage{graphicx}
\usepackage{amsfonts}
\usepackage{amsthm}
\usepackage{amsmath}
\usepackage[top=3cm, bottom=3.25cm, left=2.75cm, right=2.75cm]{geometry}
\usepackage{hyperref}

\newcommand{\beq}{\begin{equation}}
\newcommand{\eeq}[1]{\label{#1}\end{equation}}
\newcommand{\ber}{\begin{eqnarray}}
\newcommand{\eer}[1]{\label{#1}\end{eqnarray}}
\newcommand{\re}[1]{(\ref{#1})}

\newcommand{\nll}{N\!=\!(1,1)}
\newcommand{\nZZ}{N\!=\!(2,2)}
\newcommand{\nff}{N\!=\!(4,4)}

\newcommand{\ep}[1]{\epsilon^{#1}}
\newcommand{\epb}[1]{\bar{\epsilon}^{#1}}

\newcommand{\Jp}{J^{\mbox{\tiny$(+)$}}}
\newcommand{\Jm}{J^{\mbox{\tiny$(-)$}}}
\newcommand{\Jpm}{J^{\mbox{\tiny$(\pm)$}}}
\newcommand{\Up}{U^{\mbox{\tiny$(+)$}}}
\newcommand{\Um}{U^{\mbox{\tiny$(-)$}}}

\newcommand{\Upm}{U^{\mbox{\tiny$(\pm)$}}}
\newcommand{\Vpm}{V^{\mbox{\tiny$(\pm)$}}}

\newcommand{\bbD}[1]{\mathbb{D}_{#1}}
\newcommand{\bbDB}[1]{\bar{\mathbb{D}}_{#1}}
\newcommand{\bbG}[1]{\mathbb{G}_{#1}}

\newcommand{\bbX}[1]{\mathbb{X}_{#1}}
\newcommand{\bbXB}[1]{\bar{\mathbb{X}}_{#1}}

\newcommand{\bbnab}{\mbox{\hbox{$\nabla$\kern-0.65em\lower.39ex\hbox{${}^{\nabla}$}}}}

\def\+{{+\!\!\!+}}
\def\pp{\mbox{\tiny${}_{\stackrel\+ =}$}}

\newcommand{\nn}{\nonumber}
\newcommand{\be}{\begin{equation}}
\newcommand{\ee}{\end{equation}}

\numberwithin{equation}{section}

\begin{document}

\renewcommand{\theequation}{\thesection.\arabic{equation}}
\setcounter{page}{0}
\thispagestyle{empty}
\begin{flushright} \small
UUITP-16/12  
\end{flushright}
\vspace{1cm}
\begin{center}
{\LARGE $\nff$ supersymmetry and T-duality}\\[12mm]
{\large Malin G\"oteman}\\[8mm]
{\small Department of Physics and Astronomy, \\ Division of Theoretical Physics, \\ Uppsala University, \\ Box 516,
SE-751 20 Uppsala, Sweden}
\end{center}
\vspace{2cm}
\begin{abstract}
	A sigma model with four-dimensional target space parametrized by chiral and twisted chiral $\nZZ$ superfields can be extended to $\nff$ supersymmetry off-shell \cite{Gates:1984nk}, but this is not true for a model of semichiral fields, where the $\nff$ supersymmetry can only be realized on-shell \cite{Goteman:2009xb}-\cite{Goteman:2012qk}. The two models can be related to each other by T-duality. 
	
	In this paper we perform a duality transformation from a chiral and twisted chiral model with off-shell $\nff$ supersymmetry to a semichiral model. We find that additional non-linear terms must be added to the original transformations to obtain a semichiral model with $\nff$ supersymmetry, and that the algebra closes on-shell as a direct consequence of the T-duality.
\end{abstract}
\eject
%------------------------------------------------------------------------------------------------------------------------------------
\section{Introduction}

It is a well-known fact that non-linear sigma models with extended supersymmetry have constrained target space geometries. A two-dimensional $\nll$ supersymmetric sigma model without an anti-symmetric tensor in the target space allows for $\nZZ$ and $\nff$ supersymmetry if and only if the target manifold is K\"ahler and hyperk\"ahler, respectively \cite{Zumino:1979et}, \cite{AlvarezGaume:1981hm}. This interplay between supersymmetry, sigma models and geometry has  been used as a tool to investigate certain geometries; for example, generalized K\"ahler geometry can be described locally by a manifest $\nZZ$ sigma model with chiral, twisted chiral and semichiral superfields \cite{Lindstrom:2005zr}.

Supersymmetric sigma models possess a rich variety of dualities that relate different superfields and geometries \cite{Lindstrom:1983rt}-\cite{Grisaru:1997ep}. When an isometry is present in the sigma model, the gauged isometry can be used to perform a duality transformation.

A sigma model parametrized by $\nZZ$ manifestly supersymmetric chiral and twisted chiral fields allows for additional off-shell $\nff$ supersymmetry if and only if the Lagrangian satisfies a Laplace equation \cite{Gates:1984nk}, as will be reviewed in section \ref{section:(4,4)_for_chiraltwisted}. If this model possesses an abelian translational isometry, it can be dualized to a sigma model parametrized by semichiral fields describing a hyperk\"ahler manifold \cite{Bogaerts:1999jc}. 

But a sigma model parametrized by semichiral superfields cannot incorporate off-shell $\nff$ supersymmetry if the target space is four-dimensional \cite{Goteman:2009xb}. The extended supersymmetry can only occur on-shell or if the target space is $4d$-dimensional with $d>1$ \cite{Goteman:2009ye}. The explicit structure of on-shell $\nff$ supersymmetry in four dimensions was investigated in a recent paper \cite{Goteman:2012qk}.

In this paper we investigate the nature of the $\nff$ supersymmetry under T-duality. 
Symmetries that do not commute with the Killing vector field needed for T-duality transformations can not be manifest in the dual model, and rotational Killing vector fields are not compatible with abelian T-duality \cite{Alvarez:1995ai}, \cite{Bakas:1995hc}. But the Killing vector needed for the duality between a sigma model parametrized by chiral and twisted chiral fields, and one parametrized by semichiral fields, is translational and has constant components, so the isometry trivially commutes with the supersymmetry transformations. 
In general, the geometry of a dual model of a $\nff$ chiral and twisted chiral model is not necessarily hyperk\"ahler \cite{Crichigno:2011aa}, but if the isometry is translational, it is. Hence, we expect the dual semichiral model to be hyperk\"ahler.

Starting from a sigma model with $\nff$ supersymmetry, described by chiral and twisted chiral fields, we dualize to a model described by semichiral fields and analyze the obstructions of the additional supersymmetry on the dual model. We find that new non-linear terms must be added to the the known \cite{Gates:1984nk} linear transformations for the chiral and twisted chiral fields, in order to dualize into supersymmetry transformations for the semichiral model. These terms vanish when chiral and twisted chiral constraints are imposed, but prove to be necessary when performing the Legendre transformation to obtain the dual semichiral model. 

The same transformations for the semichiral model as in the recent paper \cite{Goteman:2012qk} are obtained, but the partial differential equations governing the transformations take a more transparent form in the dual framework, and the on-shell closure of the semichiral supersymmetry algebra follows directly from the T-duality procedure.

The outline of the paper is as follows: In the next section, the preliminaries regarding $\nZZ$ supersymmetric sigma models will be reviewed, and the notation will be set. Section \ref{section:relating_eqom_&_bianchi} deals with the duality of field equations and Bianchi identities under T-duality. The duality will be discussed for a simple example of a bosonic sigma model.
Section \ref{section:(4,4)_for_chiraltwisted} treats extended $\nff$ supersymmetry on sigma models and connects to results in \cite{Gates:1984nk} and \cite{Goteman:2012qk}.
In section \ref{section:tduality}, a duality transformation between a chiral and twisted chiral model, and a semichiral model will be performed and discussed in detail, and constraints on Killing vectors preserving additional supersymmetry will be derived. Section \ref{susychapter} discusses the supersymmetry transformations of the two dual models and contains the main results of the paper. The models are reduced to $\nll$ superspace in section \ref{section:reduction} and two examples are given in section \ref{section:examples}.

%------------------------------------------------------------------------------------------------------------------------------------
\section{Preface}
Consider a non-linear sigma model in $\nZZ$ superspace, described by the action
\beq
	S= \int d^2\xi d^2\theta d^2\bar\theta \,K(\phi, \bar\phi, \chi, \bar\chi, \bbX{L}, \bbXB{L}, \bbX{R}, \bbXB{R}).
\eeq{sigmamodel}
The generalized K\"ahler potential $K$ is a function of chiral $\phi$, twisted chiral $\chi$ and left/right semichiral superfields $\bbX{L,R}$ and their complex conjugates. The superfields are defined by the constraints
\ber
	\bbDB{+}\phi = \bbDB{-}\phi &=& 0,\nn\\
	\bbDB{+} \chi = \bbD{-} \chi &=&0,\nn\\
	\bbDB{+}\bbX{L} = 
	\bbDB{-}\bbX{R} &=& 0,
\eer{superfieldconstraints}
together with their complex conjugates. The covariant derivatives define the $\nZZ$ supersymmetry algebra as
\beq
	\{ \bbD{\pm}, \bbDB{\pm} \} = i\partial{\pp}.
\eeq{susyalgebra}
Whereas the chiral and the twisted chiral fields are constrained in both chiralities, the semichiral fields have only one differential constraint. This implies, for the chiral and twisted chiral fields, that half of the 16 original components fields are constrained to vanish, and another four are not independent. In total, the chiral and the twisted chiral fields depend on only four component fields, or one single $N=(1,1)$ superfield $\varphi(x,\theta^{\pm}_{\mbox{\tiny$1$}})$, 
\beq	
	\phi(x,\theta^{\pm}_{\mbox{\tiny$1$}}, \theta^{\pm}_{\mbox{\tiny$2$}}) = \varphi + \theta^{+}_{\mbox{\tiny$2$}}iD_+\varphi + \theta^{-}_{\mbox{\tiny$2$}}iD_- \varphi + \theta^{+}_{\mbox{\tiny$2$}}\theta^{-}_{\mbox{\tiny$2$}}D_+D_-\varphi,
\eeq{N=2c_hiral_in_terms_of_N=1_chiral}
and analogously for the twisted chiral fields. The semichiral fields, on the other hand, depend on two different $N=(1,1)$ superfields:
\ber
	\bbX{L} &=& X_L + \theta^{+}_{\mbox{\tiny$2$}}iD_+X_L + \theta^{-}_{\mbox{\tiny$2$}} \psi_{L\mbox{\tiny$-$}} - \theta^{+}_{\mbox{\tiny$2$}}\theta^{-}_{\mbox{\tiny$2$}}iD_+\psi_{L\mbox{\tiny$-$}},
\eer{N=2_semichiral_in_terms_of_N=1}
where $X_L(x,\theta^{\pm}_{\mbox{\tiny$1$}})$ is a bosonic superfield and $\psi_{L\mbox{\tiny$-$}}(x,\theta^{\pm}_{\mbox{\tiny$1$}})$ a fermionic. The same is valid for the right semichiral superfield $\bbX{R}$. 

The target space of the manifest $\nZZ$ supersymmetric sigma model \re{sigmamodel} is generalized K\"ahler \cite{Gualtieri:2003dx}, or bihermitian. This geometry is defined by two complex structures $\Jpm$, a metric $g$ hermitian with respect to the complex structures and an anti-symmetric $b$-field. The complex structures are covariantly constant with respect to a connection with torsion,
\beq
	\nabla^{(\pm)}\Jpm = 0, \quad \nabla^{(\pm)}=\nabla \pm \tfrac{1}{2}g^{-1} db.
\eeq{covconstant}
Whereas the chiral and twisted chiral fields parametrize the region where the two complex structures commute \cite{Ivanov:1994ec} (bilinear product space, or BiLP-geometry), the semichiral fields parametrize the region where they don't. In the semichiral section, the metric and $b$-field can be expressed in terms of the complex structures as \cite{Bogaerts:1999jc}
\beq
	g=\tfrac{1}{2}\Omega [\Jp, \Jm], \quad b = \tfrac{1}{2}\Omega \{\Jp, \Jm\},
\eeq{metricsemis}
where $\Omega$ is a certain symplectic structure. If further the target space is four-dimensional, the complex structures anti-commute to a scalar function $c$ times the identity \cite{Buscher:1987uw},
\beq
	\{\Jp, \Jm\} = 2c \cdot 1.
\eeq{Jpmanticommutator}
Reducing the sigma model to $\nll$ formalism, the complex structures can be identified in terms of second order derivatives of $K$, and the expression \re{Jpmanticommutator} can then be rewritten as
\beq
	(1+c)|K_{LR}|^2+(1-c)|K_{L\bar R}|^2=2K_{L\bar L}K_{R\bar R},
\eeq{MongeAmpere}
where the indices denote derivatives with respect to the semichiral fields. When $c$ is a constant and $|c|<1$, the torsion vanishes and the manifold is hyperk\"ahler \cite{Lindstrom:2005zr}. For $c=0$, this is equivalent to the Monge-Amp\`ere equation \cite{Sevrin:1996jr}.

In this paper, a chiral and twisted chiral sigma model will be dualized along a translational isometry to obtain a semichiral model, thus obtaining a duality between a BiLP- and hyperk\"ahler geometry. 
Semichiral sigma models were first studied in \cite{Buscher:1987uw} and further explored in several works, eg., \cite{Bogaerts:1999jc}, \cite{Sevrin:1996jr}, \cite{Lindstrom:1994mw}.
Recently, potentials for semichiral models describing hyperk\"ahler manifolds have been constructed using quotient \cite{Crichigno:2011aa} and twistor techniques \cite{Dyckmanns:2011ts}.
 
%------------------------------------------------------------------------------------------------------------------------------------
\section{Field equations and Bianchi identities}
\label{section:relating_eqom_&_bianchi}
We review here how field equations and Bianchi identities for two sigma models can be related to each other by T-duality. 
For a more extensive review, see, eg., \cite{Lindstrom:1983rt} and \cite{Hjelmeland:1997eg}.

Consider a bosonic sigma model with Euclidean target space metric and no anti-symmetric $b$-field,
\beq
	S= \int d^2\xi\, \partial_a \phi \cdot \partial^a \phi.
\eeq{bosonic_starting_model}
The coordinates of the two-dimensional world-sheet are $\xi^a$, where $a=1,2$. The action is invariant under a constant shift, $\phi \rightarrow \phi+s$. Now gauge this isometry, $\delta\phi=s(\phi)$.
In order to ensure invariance of the action, a covariant derivative must be introduced, containing a  gauge potential $V_a$ with the correct transformation properties,
\beq
	\nabla_a \phi = \partial_a \phi + V_a, \quad \delta V_a = -\partial_a s.
\eeq{cov_derivative}
The gauged action is invariant under the gauged isometry 
\beq
	S_g = \int d^2\xi \nabla_a \phi \nabla^a\phi.
\eeq{gauged_action_bosonic}
A gauge invariant field strength can be constructed as $W_{ab}=\partial_{[a} V_{b]}$. By introducing an unconstrained Lagrange multiplier $\Lambda^{ab}$, a first order action takes the form
\beq
	S_{1st} = \int d^2\xi\, \bigr[ V_aV^a+ \Lambda^{ab}W_{ab}\bigr]
	= \int d^2 \xi\, \bigr[ V_aV^a -\partial_a \Lambda^{[ab]}V_{b} \bigr],
\eeq{1st_order_action_bosonic}
where in the second step integration by parts was performed. 

Now varying this action with respect to the Lagrange multipliers gives a pure gauge condition $W_{ab}= \partial_{[a}V_{b]}=0$, which is solved by $V_a = \partial_a \phi$. Inserting this solution reproduces the original sigma model \re{bosonic_starting_model}.
The field equations for this model are simply 
\beq
0=\partial_a\partial^a\phi = \partial_a V^a. 
\eeq{eqom_bosonic_original}
Defining the Hodge star operation in terms of the totally antisymmetric tensor $\epsilon^{ab}$, the Bianchi identities are
\beq
	0 = \partial_a {}^\ast V^a = \partial _a \epsilon^{ab} V_b = \partial_{[a}V_{b]}.
\eeq{bianchi_bosonic_original}
It is clear from the form of the potential, $V_a = \partial_a \phi$, that the Bianchi identities are automatically satisfied, since partial differentials commute.

Instead varying the first order action with respect to the gauge potentials $V_a$, the resulting equation is 
\beq
	V^a = \frac{1}{2} \partial_b \Lambda^{[ba]},
\eeq{varying_bosonic_potential}
which inserted into the action gives the dual action,
\beq
	\tilde S = -\frac{1}{4}\int d^2 \xi\, \partial^b \Lambda_{[ba]} \partial_c \Lambda^{[ca]}
	= \int d^2\xi\, \partial_{[b} V_{a]} \Lambda^{ba}.
\eeq{bosonic_dual}
Varying this action with respect to $\Lambda^{ab}$ gives the field equations for the dual model,
\beq
	0=\partial_{[a}V_{b]}.
\eeq{eqom_dual_bosonic}
The Bianchi identities for the dual model are
\beq
	0 = \partial_a {}^\ast V^a = \partial_a \left(\epsilon^{ab} \epsilon_{}\frac{1}{2} \partial^c\Lambda_{[cb]}
\right) = \partial_a\left( \frac{1}{2}\partial_b \Lambda^{[ba]}\right) = \partial_a V^a.
\eeq{bianchi_bosonic_dual}
Again, from the expression of the poential in \re{varying_bosonic_potential}, one sees that the Bianchi identities are automatically fulfilled, since partial derivatives commute.

Hence, we see that the field equations for the original model, \re{eqom_bosonic_original}, takes the same form as the Bianchi identities for the dual model, \re{bianchi_bosonic_dual}, and that the Bianchi identities for the original model \re{bianchi_bosonic_original} are dual to the field equations for the dual model, \re{eqom_dual_bosonic}. To summarize, the dual models are related by
\begin{center}
\renewcommand{\arraystretch}{1.2}
\begin{tabular}{|lcl|}
	\hline
	$S=\int (\partial \phi)^2= \int V^2$&$ \longleftrightarrow$	&$\tilde S=\int (\partial \Lambda)^2= \int V^2$ \\
	field equations &  $\partial_a V^a=0$ 	& Bianchi identities\\
	Bianchi identities & $\partial_{[a}V_{b]}=0$ & field equations.\\
	\hline
\end{tabular}
\end{center}
These relations will be generalized and studied for a sigma model written in terms of manifestly $\nZZ$ superfields in section \ref{section:tduality}.

%------------------------------------------------------------------------------------------------------------------------------------
\section{The $\nff$ ``paradox"}
\label{section:(4,4)_for_chiraltwisted}
It is a well-known fact that a sigma model written in terms of one set of (anti-)chiral and twisted (anti-)chiral $\nZZ$ superfields, 
\beq
S= \int d^2\xi d^2\theta d^2\bar\theta K(\phi, \bar\phi, \chi, \bar\chi),
\eeq{model_chiralandtwisted}
can be extended to $\nff$ supersymmetry if the generalized K\"ahler potential $K$ satisfies the Laplace equation \cite{Gates:1984nk},
\beq	
	K_{\phi\bar\phi} +K_{\chi\bar\chi} = 0.
\eeq{laplace}
The $\nff$ supersymmetry transformations in four-dimensional target manifold are \cite{Gates:1984nk}
\ber
	\delta \phi &=& \epb{+}\bbDB{+}\bar\chi +\epb{-}\bbDB{-}\chi,\nn\\
	\delta \bar\phi &=& \ep{+}\bbD{+}\chi +\ep{-}\bbD{-}\bar\chi,\nn\\
	\delta \chi &=& -\epb{+}\bbDB{+}\bar\phi -\ep{-}\bbD{-}\phi,\nn\\
	\delta \bar\chi &=& -\ep{+}\bbD{+}\phi -\epb{-}\bbDB{-}\bar\phi.
\eer{simplest_4,4_ansatz}
The transformations close to a supersymmetry,
\beq
	[\delta(\epsilon_1), \delta(\epsilon_2)]X = \bar{\epsilon}_{[2}\epsilon_{1]}i\partial X
\eeq{susyclosure}
for all fields $X=(\phi, \bar\phi, \chi, \bar\chi)$, and the action \re{model_chiralandtwisted} is invariant if the Laplace equation is satisfied.

The situation completely changes for a sigma model parametrized by semichiral fields,
\beq
S= \int d^2\xi d^2\theta d^2\bar\theta K(\bbX{L}, \bbXB{L}, \bbX{R}, \bbXB{R}).
\eeq{semichiral_action}
Due to the chirality constraints on the semichiral fields, we can never find an ansatz for the semichiral fields in four-dimensional target space that closes to a $\nff$ supersymmetry off-shell.\footnote{At least for an ansatz without central charges.} Instead, an ansatz similar to \re{simplest_4,4_ansatz}, the supersymmetry of the chiral and twisted chiral superfields, can only close to a \emph{pseudo}-supersymmetry \cite{Goteman:2009xb},
\beq
	[\delta(\epsilon_1), \delta(\epsilon_2)]\bbX{} = -\bar{\epsilon}_{[2}\epsilon_{1]}i\partial \bbX{}.
\eeq{pseudosusy}
In four dimensions, the supersymmetry is realized only on-shell; only when the target space is enlarged to $4d$ dimensions with $d>1$, an ansatz can be written down that closes to $\nff$ supersymmetry off-shell \cite{Goteman:2009ye}.

Transformations that close to an on-shell $\nff$ supersymmetry for a set of semichiral fields can be written as \cite{Goteman:2012qk}
\ber\nn
\delta \bbX{L} &=&\bar\epsilon^+ \bbDB{+}f(\bbX{L},\bbXB{L},\bbX{R},\bbXB{R}) + \bar\epsilon^{-}\bbDB{-} \bbX{L} - \epsilon ^- \bbD{-}\bbX{L},\\
\delta \bbXB{L} &=& \ep{+}\bbD{+} \bar f(\bbX{L},\bbXB{L},\bbX{R},\bbXB{R}) + \ep{-}\bbD{-}\bbXB{L} - \epb{-}\bbDB{-}\bbXB{L},\nn\\
\delta \bbX{R}&=& \bar\epsilon^- \bbDB{-}\hat f(\bbX{L},\bbXB{L},\bbX{R},\bbXB{R}) -\bar\epsilon^+ \bbDB{+}\bbX{R}+\epsilon^+ \bbD{+}\bbX{R},\nn\\
\delta\bbXB{R} &=& \ep{-}\bbD{-} \bar{\hat f}(\bbX{L},\bbXB{L},\bbX{R},\bbXB{R}) -\ep{+}\bbD{+}\bbXB{R}+ \epb{+}\bbDB{+}\bbXB{R}.
\eer{defining_ansatz_2}
Invariance of the action relates the derivatives of the functions $f$ and $\hat f$, implying that the algebra closes when the field equations are imposed. 

If the actions possesses a certain abelian isometry, a chiral and twisted chiral model can be dualized into a semichiral one. It is then an interesting question to ask what happens to the off-shell $\nff$ supersymmetry in the chiral and twisted chiral sigma model, when dualized to a semichiral model, and if the transformations for the semichiral fields \re{defining_ansatz_2} can be related to the transformations of the chiral and twisted chiral fields \re{simplest_4,4_ansatz}. These questions will be explored in the next sections.

%------------------------------------------------------------------------------------------------------------------------------------
\section{T-duality between $(\phi, \chi)$ and $(\bbX{L}, \bbX{R})$}
\label{section:tduality}
\subsection{Duality transformation}
Consider a sigma model with chiral and twisted chiral superfields $K(\phi, \chi)$ and an isometry defined by the Killing vector
\beq
	k=i(\partial_\phi-\partial_{\bar\phi}-\partial_\chi+\partial_{\bar\chi}).
\eeq{killing_chiraltwisted}
In combinations of coordinates adapted to this isometry, the action is of the form
\beq
K= K\left(\phi+\bar\phi, \chi+\bar \chi, i(\phi-\bar\phi+\chi-\bar \chi) \right).
\eeq{chiral_twisted_isometry}
Under the gauged isometry, the fields transform as
\beq
	\delta\phi = i\Lambda, \quad \delta \chi = i\tilde\Lambda
\eeq{global_shifts_chiraltwisted}
together with their complex conjugates, where $\Lambda$ and $\tilde\Lambda$ satisfy chiral and twisted chiral chirality properties, respectively.
In order to keep invariance of the action under the gauged isometry, gauge potentials must be introduced. The gauge potentials are real and transform as
\beq
	\delta V_\phi = -i(\Lambda-\bar\Lambda), \quad 
	\delta V_\chi = -i(\tilde\Lambda-\bar{\tilde\Lambda}), \quad 
	\delta V = \Lambda+\bar\Lambda+\tilde\Lambda+\bar{\tilde\Lambda}.
\eeq{Vl_V'_potentials}
This is the so called large vector multiplet (LVM) \cite{Lindstrom:2007vc}. In \cite{arXiv:1008.3186}, it was shown that the large vector multiplet does not allow for off-shell $\nff$ supersymmetry. The gauged action takes the form
\beq
	S_{\mbox{\tiny gauged}}= \int d^2\xi d^2\theta d^2\bar\theta K\left(\phi+\bar\phi+V_\phi, \chi+\bar\chi+{V}_\chi, i(\phi-\bar\phi+\chi-\bar\chi)+V\right).
\eeq{gauged_action_Vl_V'}
Gauge invariant field strengths can be defined in terms of the gauge potentials as
\beq
	W=i\bbD{-}\bbDB{+}V_\phi, \quad \tilde W= i\bbDB{-}\bbDB{+}V_\chi
\eeq{higherorder_fieldstrengths}
together with their complex conjugates. The field strengths $W$ and $\tilde W$ satisfy twisted chiral and chiral constraints, respectively.
To perform a T-duality transformation to a sigma model parametrized by semichiral fields, we will need to introduce field strengths satisfying semichiral constraints. To do this, define two new complex gauge potentials as
\ber
	V_L = \frac{1}{2}\bigr(V+i(V_\phi+V_\chi)\bigr),\quad& \delta V_L = \Lambda+\tilde \Lambda \nn\\
	V_R = \frac{1}{2}\bigr(V+i(V_\phi-V_\chi)\bigr), \quad & \delta V_R = \Lambda+\bar{\tilde \Lambda}.
\eer{VLVR}
The complex gauge potentials $V_L$ and $V_R$ transform into left- and right semichiral gauge parameters, respectively, and they satisfy the reality condition
\beq
	V_L+\bar V_L = V_R + \bar V_R.
\eeq{Vlreal=Vrreal}
From the transformation properties of the gauge potentials, it is clear that gauge invariant field strengths can be constructed as
\beq
	\mathbb{G}_+ = \bbDB{+}V_L, \quad \mathbb{G}_-= \bbDB{-} V_R.
\eeq{semichiral_fieldstrengths}

Introducing spinorial Lagrange multipliers $X^\pm$ and choosing gauge such that $\phi+\bar\phi=0$, $\chi+\bar\chi=0$ and $i(\phi-\bar\phi+\chi-\bar\chi)=0$, a first order action can be defined in terms of the gauge potentials as
\ber
	S_{\mbox{\tiny 1st}}&=&\int d^2\xi d^2\theta d^2\bar\theta \Bigr[ K(V_\phi, V_\chi, V)- (X^+\mathbb{G}_+ + \bar{X}^+\bar{\mathbb{G}}_+ + X^-\mathbb{G}_- +\bar{X}^-\bar{\mathbb{G}}_-)\Bigr]\nn\\
	&=&\int d^2\xi d^2\theta d^2\bar\theta \Bigr[ K\bigr(-\tfrac{i}{2}(V_L-\bar{V}_L+V_R-\bar{V}_R), -\tfrac{i}{2}(V_L-\bar{V}_L-V_R+\bar{V}_R), V_L+\bar{V}_L\bigr)\nn\\
	&&\phantom{\int d^2\xi d^2\theta d^2\bar\theta \Bigr[}- (X^+\mathbb{G}_+ + \bar{X}^+\bar{\mathbb{G}}_+ + X^-\mathbb{G}_- +\bar{X}^-\bar{\mathbb{G}}_-)\Bigr].
\eer{firstorderactionfirst}
Varying this action with respect to the Lagrange multipliers implies that the field strengths vanish, $\bbG{+}=\bbG{-}=0$. This pure gauge condition together with the reality constraint \re{Vlreal=Vrreal} imply that
\beq
	V_L = i(\phi+\chi), \quad V_R =i(\phi-\bar\chi),
\eeq{puregauge}
or, equivalently, 
\beq
V_\phi=\phi+\bar\phi,\quad
V_\chi = \chi+\bar\chi,\quad
V=i(\phi-\bar\phi+\chi-\bar\chi).
\eeq{puregaugeVphietc}
For clarity we now rename the coordinates as $V_\phi=x$, $V_\chi=y$ and $V=z$, and similarly 
\beq
x_L=V_L = \frac{1}{2}(z+ix+iy), \quad
x_R=V_R = \frac{1}{2}(z+ix-iy).
\eeq{defining_XLR}
Inserting this into the first order action, the orginal action \re{chiral_twisted_isometry} is recovered,
\beq
S_{\mbox{\tiny original}} = \int d^2\xi d^2\theta d^2\bar\theta K\bigr(\underbrace{\phi+\bar\phi}_{x}, \underbrace{\chi+\bar\chi}_{y}, \underbrace{i(\phi-\bar\phi+\chi-\bar\chi)}_{z}\,\bigr).
\eeq{original_again}

On the other hand, integrating the first order action \re{firstorderactionfirst} by parts, new semichiral fields can be defined as $\bbDB{+}X^+= \bbX{L}$ and $\bbDB{-}X^-=\bbX{R}$. Using the reality constraint \re{Vlreal=Vrreal}, the first order action takes the form
\beq
	S_{\mbox{\tiny 1st}}=\int d^2\xi d^2\theta d^2\bar\theta \bigr[ K(x, y, z)- \tilde{x}x-\tilde{y}y-\tilde{z}z\bigr],
\eeq{firstorder_part_ingegrated}
where we defined combinations of the semichiral fields as
\ber
	\tilde{x} &=& \tfrac{i}{2}(\bbX{L}-\bbXB{L}+\bbX{R}-\bbXB{R}),\nn\\
	\tilde{y} &=& \tfrac{i}{2}(\bbX{L}-\bbXB{L}-\bbX{R}+\bbXB{R}),\nn\\
	\tilde{z} &=& \tfrac{1}{2}(\bbX{L}+\bbXB{L}+\bbX{R}+\bbXB{R}).
\eer{xyz_def}
Varying this action now with respect to the gauge potentials gives
\beq
	\frac{\partial K}{\partial x} = \tilde{x},\quad
	\frac{\partial K}{\partial y} = \tilde{y}, \quad
	\frac{\partial K}{\partial z} = \tilde{z}
\eeq{varyingpotentials} 
which in turn implies
\beq
x=x(\tilde{x}^i),\quad y=y(\tilde{x}^i),\quad z=z(\tilde{x}^i),
\eeq{potentials_of_xyz}
where by $\tilde{x}^i$ we denote the three dual coordinates, $\tilde{x}^i=(\tilde{x},\tilde{y},\tilde{z})$.
Inserting these relations into the first order action \re{firstorder_part_ingegrated} gives the dual model,
\ber
S_{\mbox{\tiny dual}} &=& \int d^2\xi d^2\theta d^2\bar\theta \Bigr[K\bigr(x(\tilde{x}^i), y(\tilde{x}^i), z(\tilde{x}^i)\bigr)-\tilde{x}x(\tilde{x}^i)-\tilde{y}y(\tilde{x}^i)-\tilde{z}z(\tilde{x}^i)\Bigr]\nn\\
 &=& \int d^2 \xi d^2\theta d^2\bar\theta\; \tilde K(\tilde{x},\tilde{y},\tilde{z}).
\eer{dualmodel}
This duality procedure shows the equivalence between a sigma model written in terms of chiral and twisted chiral fields, \re{chiral_twisted_isometry}, with one written in terms of semichiral fields, \re{dualmodel}. The dual model also possesses an abelian isometry, given by the fields $\tilde{x}$, $\tilde{y}$ and $\tilde{z}$ in \re{xyz_def}.
The fourth independent coordinate, parametrizing the isometry direction, is $\tilde w=\frac{1}{2}(\bbX{L}+\bbXB{L}-\bbX{R}-\bbXB{R})$. The chirality constraints of the semichiral fields imply the relation between $\tilde w$ and the other three coordinates $\tilde{x}^i$ as $\bbDB{\pm} \tilde{w} = \bbDB{\pm}(\pm i\tilde{x}+i\tilde{y} \mp \tilde{z})$.

%_________________________________________________________________________________________
\subsection{Killing vectors preserving $\nff$}
\label{section: Killing_vectors}
For a T-duality with Killing vector $k$ to preserve $\nff$ supersymmetry of the twisted multiplet \cite{Gates:1984nk}, it must preserve the complex structures $\Jpm_1$, $\Jpm_2$ generating the supersymmetry transformations, i.e.,
\beq
\mathcal{L}_{k}J^{\mu}_{\,\,\,\nu}= k^{\rho} \partial_{\rho}J^{\mu}_{\,\,\, \nu} -  \partial_{\rho}k^{\mu} J^{\rho}_{\,\,\, \nu}+\partial_{\nu}k^{\rho} J^{\mu}_{\,\,\, \rho}= 0.
\eeq{vanishingKilling}
In the coordinates $(\phi, \bar\phi, \chi, \bar\chi)$ describing the $\nff$ twisted multiplet, the complex structures are constant \cite{Gates:1984nk} and the first term vanishes. Therefore, $\partial_{\rho}k^{\mu} J^{\rho}_{\,\,\, \nu}= J^{\mu}_{\,\,\, \rho}\partial_{\nu}k^{\rho}$
needs to be satisfied. 
This relation implies that the Killing vector is of the form
\beq
k=k^{\phi}(\phi)\partial_{\phi}+k^{\bar \phi}(\bar \phi)\partial_{\bar \phi}+k^{\chi}(\chi)\partial_{\chi}+k^{\bar \chi}(\bar \chi)\partial_{\bar \chi},
\eeq{killingvectoreqn}
so that the matrix $\partial_{\mu}k^{\nu}$ is diagonal, 
\beq
\partial_{\mu}k^{\nu}=\left(
\begin{array}{cccc}
\frac{\partial k^{\phi}}{\partial \phi}& 0 & 0 & 0  \\ 
0 & \frac{\partial k^{\bar \phi}}{\partial \bar \phi}& 0 & 0\\
0 & 0 & \frac{\partial k^{\chi}}{\partial \chi} & 0  \\ 
 0 & 0 & 0 & \frac{\partial k^{\bar \chi}}{\partial \bar \chi}
\end{array}\right).
\eeq{diagonalkilling}
Further, the relation $\partial_{\rho}k^{\mu} J^{\rho}_{\,\,\, \nu}= J^{\mu}_{\,\,\, \rho}\partial_{\nu}k^{\rho}$ implies, for the complex structure $\Jp_1$, that $k^\phi_{,\phi}=k^\chi_{,\chi}$, and for the complex structure $\Jm_1$ that $k^\phi_{,\phi}=k^{\bar\chi}_{,\bar\chi}$.\footnote{
The author wishes to thank Marcos Crichigno and Martin Ro\v{c}ek for developing this idea.} The complex structures $\Jpm_2$ give the same conditions. 
In total, the coefficients of the Killing vector must satisfy the constraints
\beq
\frac{\partial k^{\phi}}{\partial \phi}=\frac{\partial k^{\bar \phi}}{\partial \bar \phi}=\frac{\partial k^{\chi}}{\partial \chi}=\frac{\partial k^{\bar \chi}}{\partial \bar \chi}.
\eeq{killing_must_satisfy}
The only solutions to this are either that all components in the Killing vector are constants, $\partial_{\mu}k^{\nu}=0$, or that they are linear with the same coefficient. 
That is, the only isometries preserving the $\nff$ supersymmetry are
\ber
\text{translations:}\quad k &=&i(\partial_{\phi}-\partial_{\bar \phi}-\partial_{\chi}+\partial_{\bar \chi}), \\ 
\text{rescalings:}\quad k &=&\phi \partial_{\phi}+ \bar \phi \partial_{\bar \phi}+\chi \partial_{\chi}+ \bar \chi \partial_{\bar \chi}.
\eer{isometriespreserving(4,4)}

In the previous section \ref{section:(4,4)_for_chiraltwisted}, we have seen that a sigma model written in terms of chiral and twisted chiral fields can be extended to off-shell $\nff$ supersymmetry with the linear supersymmetry transformations in \re{simplest_4,4_ansatz}, but that the analogue transformations \re{defining_ansatz_2} for the semichiral fields are non-linear and can only close to a supersymmetry on-shell. 
Starting from a sigma model parametrized by chiral and twisted chiral fields and having $\nff$ supersymmetry, we can find the dual model parametrized by semichiral fields by the method discussed in this section. We have also seen that the translational isometries needed for the duality should preserve the $\nff$ supersymmetry. The questions are: what happens to the $\nff$ supersymmetry from the original model; can we write the non-linear supersymmetry transformations of the semichiral fields in terms of the linear ones for the chiral and twisted chiral, and how can we understand the on-shell condition?

We will in the next section show that the Bianchi identities of the original model are satisfied and correspond to field equations in the dual model, such that the dual model can accomodate $\nff$ supersymmetry on-shell. Further, we will relate the supersymmetry transformations of the dual models and show how the non-linear terms in the transformations arise.

%_________________________________________________________________________________________
\section{$\nff$ supersymmetry in the two dual models}
\label{susychapter}
%_________________________________________________________________________________________
\subsection{Supersymmetry in original model}
Consider the original model \re{original_again}. As expected, it is invariant under the supersymmetry transformations \re{simplest_4,4_ansatz} if and only if the generalized K\"ahler potential satisfies the analogue of the Laplace equation \re{laplace} for a sigma model with the isometry \re{killing_chiraltwisted},
\beq
	K_{x x}+K_{y y} + 2K_{zz} = 0.
\eeq{laplacewithisometry}

The supersymmetry transformations on the coordinates that are combinations of chiral and twisted chiral superfields,
\beq
x^\mu = (x, y, z),
\eeq{original_coordinates_first} 
can be derived from the transformations on the chiral and twisted chiral fields given in equation \re{simplest_4,4_ansatz} and read
\ber
	\delta x &=& \epb{+}\bbDB{+}y+\epb{-}\bbDB{-}y+\ep{+}\bbD{+}y+\ep{-}\bbD{-}y \nn\\
	\delta y &=& -\epb{+}\bbDB{+}x-\epb{-}\bbDB{-}x-\ep{+}\bbD{+}x-\ep{-}\bbD{-}x \label{susytransformations_original_coordinates_2}\\
	\delta z &=& i\epb{+}\bbDB{+}(y-x)+i\epb{-}\bbDB{-}(y+x)-i\ep{+}\bbD{+}(y-x)-i\ep{-}\bbD{-}(y+x).\nn
\eer{susytransformations_original_coordinates}
Note, that due to the chirality constraints of the chiral and the twisted chiral fields there is an ambiguity in the expressions, as the Bianchi identities,
\ber
\bbDB{+}x_L &=& \tfrac{1}{2}\bbDB{+}(z+ix+iy) =0, \nn\\
\bbDB{-}x_R &=& \tfrac{1}{2}\bbDB{-}(z+ix-iy) =0,
\eer{Bianchichiraltwisted}
allow us to add terms to the transformations, for example $\delta x = i\alpha\epb{+}\bbDB{+}x_L$ plus the complex conjugate, where $\alpha$ is some arbitrary function. The terms might seem unnecessary since they vanish due to the chirality constraints, but they will prove to be crucial when discussing the supersymmetry transformations of the dual semichiral model later. 
Considering this ambiguity, the most general form of the transformations on $x^\mu$ is
\beq
	\delta x^\mu = \epb{+} U^{(+)}{}^\mu_\nu \bbDB{+} x^\nu+\epb{-} U^{(-)}{}^\mu_\nu \bbDB{-} x^\nu + \ep{+} V^{(+)}{}^\mu_\nu \bbD{+} x^\nu+\ep{-} V^{(-)}{}^\mu_\nu \bbD{-} x^\nu,
\eeq{susytransformations_original_ambiguity}
where the transformation matrices take the form
\beq
	U^{(+)} = \frac{1}{2}\left(\begin{array}{ccc}
	-\alpha & (2-\alpha) & i\alpha\\
	-(2+\gamma) & -\gamma & i\gamma \\
	-i(2+\varepsilon) & i(2-\varepsilon) & -\varepsilon
	\end{array}\right), \quad
	U^{(-)} = \frac{1}{2}\left(\begin{array}{ccc}
	-\beta & (2+\beta) & i\beta\\
	-(2+\delta) & \delta & i\delta \\
	i(2-\kappa) & i(2+\kappa) & -\kappa
	\end{array}\right)
\eeq{transf_matrices}
and $\Vpm$ are the complex conjugates of $\Upm$.
The parameters $\alpha, \beta, \gamma, \delta, \varepsilon$ and $\kappa$ are arbitrary functions and will not appear in the transformations when the chirality constraints are used; the transformations thus take the well-known form as in (\ref{susytransformations_original_coordinates_2}). The transformations close to a supersymmetry,
\beq
	[\delta(\ep{}_1), \delta(\ep{}_2)]x^\mu = \epb{+}_{[2} \ep{+}_{1]}i\partial_{\+} x^\mu + \epb{-}_{[2} \ep{-}_{1]}i\partial_{=} x^\mu.
\eeq{susy_close_on_original}

%_________________________________________________________________________________________
\subsection{Field equations and Bianchi identities}
\label{section:Field equations and Bianchi identities}
The field equations of the original model \re{original_again} are obtained by varying the action with respect to the unconstrained fields. The chiral and twisted chiral fields are constrained and can be written in terms of some unconstrained fields as $\phi=\bbDB{+}\bbDB{-}\lambda$ and $\chi=\bbDB{+}\bbD{-}\tilde \lambda$. Varying the original action with respect to $\lambda$ and $\tilde \lambda$ gives the field equations
\beq
	\begin{array}{llcl}
	\delta\lambda: \quad& \bbDB{+}\bbDB{-}\bigr[K_{z}-iK_x\bigr] &=& 0,\\
	\delta\tilde\lambda: \quad& \bbDB{+}\bbD{-}\bigr[K_{z}-iK_y\bigr] &=& 0.
	\end{array}
\eeq{eqom_original}

For the dual model, the unconstrained fields are the Lagrange multipliers $X^{\pm}$. Varying the action \re{dualmodel} with respect to $X^{\pm}$ gives the equations of motion for the dual semichiral model,
\beq
	\begin{array}{llcl}
	\delta X^+: \quad& \bbDB{+}\bigr[\tilde K_{\tilde{z}}+ i\tilde K_{\tilde{x}}+i\tilde K_{\tilde{y}}\bigr] &=& 0,\\
	\delta X^-: \quad& \bbDB{-}\bigr[\tilde K_{\tilde{z}}+i \tilde K_{\tilde{x}}-i\tilde K_{\tilde{y}}\bigr] &=& 0.
	\end{array}
\eeq{eqom_dual}
In terms of the original generalized K\"ahler potential $K$, the derivatives $\tilde K_{\tilde{x}}$, $\tilde K_{\tilde{y}}$ and $\tilde K_{\tilde{z}}$ are simply derived from \re{dualmodel},
\ber
	\tilde K_{\tilde{x}} &=\ K_{x}x'(\tilde{x}^i)-x(\tilde{x}^i)-\tilde{x}x'(\tilde{x}^i) &=\ -x(\tilde{x}^i),\nn\\
	\tilde K_{\tilde{y}} &=\ K_{y}y'(\tilde{x}^i)-y(\tilde{x}^i)-\tilde{y}y'(\tilde{x}^i)&=\ -y(\tilde{x}^i),\nn\\
	\tilde K_{\tilde{z}} &=\ \ \, K_{z}z'(\tilde{x}^i)-z(\tilde{x}^i)-{\tilde{z}}z'(\tilde{x}^i)&=\ -z(\tilde{x}^i),
\eer{KtildeK_eqom}
or in short, $\tilde K_i = -\delta_{i\mu}x^\mu$.

In the discussion of T-duality for the bosonic sigma model in section \ref{section:relating_eqom_&_bianchi}, we saw that the Bianchi identities in the dual models were automatically satisfied due to the expressions of the potentials, and that the Bianchi identities in one model correspond to the field equations in the dual model, and vice versa. The same is true here. 

From the expression of the derivatives $K_{x}$, $K_{y}$ and $K_z$ in \re{xyz_def}-\re{varyingpotentials}, one finds the Bianchi identities in the dual model \re{dualmodel},
\ber
	\bbDB{+}\bbDB{-}[\tilde z-i\tilde x] = \bbDB{+}\bbDB{-}[\bbX{L}+\bbX{R}] &=&0,\nn\\
	\bbDB{+}\bbD{-}[\tilde z-i\tilde y] = \bbDB{+}\bbD{-}[\bbX{L}+\bbXB{R}] &=&0.
\eer{bianchi_dual_(2,2)}
These correspond to the field equations of the original model, \re{eqom_original}.
Similarly, using the expressions in \re{KtildeK_eqom} together with the form of the potentials in \re{varyingpotentials}, the Bianchi identities in the original model, \re{Bianchichiraltwisted}, take the form
\ber
	i\bbDB{+}[z+i(x+y)] &=& 0,\nn\\
	i\bbDB{-}[z+i(x-y)] &=& 0.
\eer{bianchi_original_(2,2)} 
These are automatically satisfied when $(x,y,z)$ are identified as the combinations of chiral and twisted chiral fields in \re{puregaugeVphietc}, and are equivalent to the field equations of the dual model, \re{eqom_dual}.

In summary, equations \re{eqom_original} are the field equations for the original model and the Bianchi identity for the dual model, where they are automatically satisfied. Analogously, \re{eqom_dual} are the equations of motion for the dual model and the Bianchi identities for the original model.

%_________________________________________________________________________________________
\subsection{Supersymmetry transformations in the dual model}
The coordinates of the original model \re{original_again} are $x^\mu=(x, y, z)$ and the coordinates of the dual model \re{dualmodel} are $\tilde{x}^i = (\tilde{x},\tilde{y},\tilde{z})$. Whereas $x^\mu$ are combinations of chiral and twisted chiral fields, $\tilde{x}^i$ are combinations of semichiral fields. The two coordinate systems are related by equation \re{varyingpotentials},
\beq
	K_\mu = \delta_{\mu i}\tilde{x}^i.
\eeq{relating_coordinates}
All models we will consider require that the $3 \times 3$ matrix $\partial_{x^\mu}\partial_{x^\nu}K=K_{\nu\mu}$ is invertible. Hence, the relation \re{relating_coordinates} implies
\beq
	\bbD{}x^\mu = (K^{-1})^{\mu\nu}\delta_{\nu i}\bbD{}\tilde{x}^i
\eeq{relating_D_on_coordinates}
as well as
\beq
	\tilde K_{ij}=-\delta_{i\mu}(K^{-1})^{\mu\nu}\delta_{\nu j}.
\eeq{relating_Ks}

With these relations and the supersymmetry transformations of the original coordinates given in \re{susytransformations_original_ambiguity}, the supersymmetry transformations of the dual coordinates can now be derived as
\ber
	\delta \tilde{x}^i &=& \delta^{i\mu}K_{\mu\nu}\delta x^\nu\nn\\
	&=& \delta^{i\mu}K_{\mu\nu}\bigr(\epb{\alpha} U^{(\alpha)}{}^\nu_\rho \bbDB{\alpha} x^\rho + \ep{\alpha} V^{(\alpha)}{}^\nu_\rho \bbD{\alpha} x^\rho\bigr)\nn\\
	&=& \epb{\alpha}\bigr(\delta^{i\mu}K_{\mu\nu} U^{(\alpha)}{}^\nu_\rho (K^{-1})^{\rho\sigma}\delta_{\sigma j}\bigr)\bbDB{\alpha}\tilde{x}^j + \ep{\alpha}\bigr(\delta^{i\mu}K_{\mu\nu} V^{(\alpha)}{}^\nu_\rho (K^{-1})^{\rho\sigma}\delta_{\sigma j}\bigr)\bbD{\alpha}\tilde{x}^j\nn\\
	&=& \epb{\alpha}\tilde{U}^{(\alpha)}{}^i_j \bbDB{\alpha}\tilde{x}^j + \ep{\alpha}\tilde V^{(\alpha)}{}^i_j \bbD{\alpha}\tilde{x}^j,
\eer{dual_susytransformations}
where $\alpha=+,-$ is the spinorial index and we defined the matrices $\tilde U^{(\alpha)}$ and $\tilde V^{(\alpha)}$ as
\ber
\tilde{U}^{(\alpha)}{}^i_j &=& \delta^{i\mu}K_{\mu\nu} U^{(\alpha)}{}^\nu_\rho (K^{-1})^{\rho\sigma}\delta_{\sigma j},\nn\\
\tilde{V}^{(\alpha)}{}^i_j &=& \delta^{i\mu}K_{\mu\nu} V^{(\alpha)}{}^\nu_\rho (K^{-1})^{\rho\sigma}\delta_{\sigma j},
\eer{dual_transformation_matrices}
where we can recall that $V^{(\alpha)}=\bar{U}^{(\alpha)}$. 

In the related paper \cite{Goteman:2012qk}, the explicit on-shell $\nff$ supersymmetry transformations have been written down for a model of semichiral fields in four-dimensional target space and take, after some parameters have been absorbed in rescalings of the fields, the expression in \re{defining_ansatz_2}. Using this result, an ansatz for the dual coordinates $\tilde x^i$ can be written down as
\ber
\delta \tilde x &=&\tfrac{i}{2}\epb{+}\bigr[\bbDB{+} f -\bbDB{+}(\tilde z-i\tilde x)-\bbDB{+}(\tilde z-i\tilde y)\bigr] +\tfrac{i}{2}\epb{-}\bigr[\bbDB{-} \hat f +\bbDB{-}(\tilde z-i\tilde x)+\bbDB{-}(\tilde z+i\tilde y)\bigr],\nn\\
\delta \tilde y &=&\tfrac{i}{2}\epb{+}\bigr[\bbDB{+} f +\bbDB{+}(\tilde z-i\tilde x)+\bbDB{+}(\tilde z-i\tilde y)\bigr]-\tfrac{i}{2}\epb{-}\bigr[\bbDB{-} \hat f -\bbDB{-}(\tilde z-i\tilde x)-\bbDB{-}(\tilde z+i\tilde y)\bigr], \nn\\
\delta \tilde z &=&\tfrac{i}{2}\epb{+}\bigr[\bbDB{+} f -\bbDB{+}(\tilde z-i\tilde x)+\bbDB{+}(\tilde z-i\tilde y)\bigr]+\tfrac{i}{2}\epb{-}\bigr[\bbDB{-} \hat f +\bbDB{-}(\tilde z-i\tilde x)-\bbDB{-}(\tilde z+i\tilde y)\bigr].\nn\\
\eer{transformations_on_semichirals}
plus the complex conjugate parts. The parameters $f=f(\tilde x^i)$ and $\hat f=\hat f(\tilde x^i)$ are both functions of the combinations of the semichiral coordinates, and
the transformations are constructed as to satisfy the Bianchi identities for the semichiral fields,
\beq
	\bbDB{+}\bbDB{-}\bigr(\delta(\tilde z-i\tilde x)\bigr) = 0, \quad 	\bbDB{+}\bbD{-}\bigr(\delta(\tilde z-i\tilde y)\bigr) = 0.
\eeq{bianchi_for_transformations}
The compact form of the transformations in \re{transformations_on_semichirals} reads
\beq
	\delta \tilde x^i= \epb{\alpha}\tilde{U}^{(\alpha)}{}^i_j \bbDB{\alpha}\tilde{x}^j + \ep{\alpha}\tilde V^{(\alpha)}{}^i_j \bbD{\alpha}\tilde{x}^j,
\eeq{transformations_semichirals_compact}
where the transformation matrices take the form
\ber
	\tilde{U}^{(+)}&=&\frac{1}{2}\left(\begin{array}{ccc} i f_{\tilde x} -1& i f_{\tilde y} -1& i (f_{\tilde z} -2)\\ 
	i f_{\tilde x} +1& i f_{\tilde y} +1& i (f_{\tilde z} +2)\\
	f_{\tilde x} +i &  f_{\tilde y} -i& f_{\tilde z}\end{array}\right),\nn\\
\tilde{U}^{(-)}&=&\frac{1}{2}\left(\begin{array}{ccc} i \hat f_{\tilde x} +1& i \hat f_{\tilde y} -1& i (\hat f_{\tilde z} +2)\\ 
	-i \hat f_{\tilde x} +1& -(i \hat f_{\tilde y} +1)& -i (\hat f_{\tilde z} -2)\\
	\hat f_{\tilde x} -i &  \hat f_{\tilde y} -i& \hat f_{\tilde z}\end{array}\right),
\eer{transformation_matrices_semis}
where the indices denote derivatives; $f_{\tilde x}= \tfrac{\partial}{\partial \tilde x}f$.
The relation between these expressions and the expressions in \re{dual_transformation_matrices} will be derived in the next subsection.

In \cite{Goteman:2012qk}, it was shown that the transformations on the semichiral fields close to a supersymmetry on-shell and the semichiral action $\int K(\bbX{})$ is invariant if and only if the transformation parameters are certain functions of second order derivatives of $K$. In the next two subsections we will show that the same is valid here, and that the on-shell algebra closure follows directly from T-duality. 

%_________________________________________________________________________________________
\subsection{Invariance of action}
\label{Invariance of action}
The original chiral and twisted chiral action \re{original_again} is invariant under the supersymmetry transformations \re{susytransformations_original_ambiguity} if the generalized K\"ahler potential satisfies $(K_\mu U^{(\alpha)}{}^\mu_{[\nu})_{\rho]}\bbDB{\alpha}x^\nu \bbDB{\alpha}x^\rho=0$. Using the Bianchi identities for the chiral and twisted chiral fields, this is proportional to the Laplace equation \re{laplacewithisometry},
\beq
	\delta S = 0 \quad \Longleftrightarrow \quad K_{xx}+ K_{yy}+2K_{zz} = 0.
\eeq{laplace_reminder}
The Legendre transformation implies that the dual potentials are related by $K=-\tilde K^{-1}$  \re{relating_Ks} , so the linear Laplace equation corresponds to the following non-linear relation for the dual potential,
\beq
(\tilde K_{\tilde{x}\tilde{x}} + \tilde K_{\tilde{y}\tilde{y}})\tilde K_{\tilde{z}\tilde{z}}  +2\tilde K_{\tilde{x}\tilde{x}}\tilde K_{\tilde{y}\tilde{y}} - 2\tilde K_{\tilde{x}\tilde{y}}^2-\tilde K_{\tilde{x}\tilde{z}}^2-\tilde K_{\tilde{y}\tilde{z}}^2=0.
\eeq{laplace_dual}
It is known from \cite{Bogaerts:1999jc} that the Laplace equation is dualized into an equation equivalent to the Monge-Amp\`ere equation, if one performs a T-duality along a translational isometry. Therefore, \re{laplace_dual} is nothing but equation \re{MongeAmpere} with $c=0$, which is  equivalent to the Monge-Amp\`ere equation, and so the dual target space is hyperk\"ahler. It is interesting to recall that one can perform a T-duality along a rescaling isometry, while still preserving $\nff$. In this case, the dual potential does not describe a hyperk\"{a}hler manifold and therefore the invariance of the original action does not correspond to the Monge-Amp\`ere equation \cite{Crichigno:2011aa}.  

Now we turn to invariance of the dual semichiral model and to the identification of the new non-linear terms in $U$. 
If the chiral and twisted chiral constraints, i.e. the Bianchi identities in the original model, were not to be used, $K_{\mu[\nu}\Upm{}^\mu_{\rho]}=0$ would imply that some of the parameters $\alpha$, $\beta$, $\gamma$, $\delta$, $\varepsilon$ and $\kappa$ in \re{transf_matrices} could be solved in terms of the others and second derivatives of $K$. The invariance of the $\Up$-transformations relate the parameters $\alpha$, $\gamma$ and $\varepsilon$ and the $\Um$-transformations relate the parameters $\beta$, $\delta$ and $\kappa$, 
\beq
	\alpha=\alpha(\partial\partial K, \varepsilon), \quad 
	\beta=\beta(\partial\partial K, \kappa), \quad 
	\gamma=\gamma(\partial\partial K, \varepsilon), \quad
	\delta=\delta(\partial\partial K, \kappa).
\eeq{solvealphabeta}
Now define the parameters to be 
\beq
	\begin{array}{lcllcl}
	\alpha &=& -i(K^{-1})^{1\mu} f_{\mu}+1,\quad &
	\beta &=& -i(K^{-1})^{1\mu} \hat f_{\mu}-1,\\
	\gamma &=& -i(K^{-1})^{2\mu} f_{\mu}-1,\quad &
	\delta &=& -i(K^{-1})^{2\mu} \hat f_{\mu}-1,\\
	\varepsilon &=& - (K^{-1})^{3\mu}f_{\mu},&
	\kappa &=& -(K^{-1})^{3\mu}\hat f_{\mu},
	\end{array}
\eeq{definingalphabeta}
for some arbitrary functions $f$ and $\hat f$, where the indices denote derivative with respect to the coordinates $x^\mu$. Defining the parameters in this way will ensure that the obtained transformations for the combinations of semichiral fields agree with the transformations obtained in \cite{Goteman:2012qk}. Hence, the parameters depend on derivatives of some function and second order derivatives of $K$, and the constraints in \re{solvealphabeta} applied to the expressions in \re{definingalphabeta} then implies that the functions $f$ and $\hat f$ must satisfy the following partial differential equations,
\ber
f_z+if_x &=& -(K_x+K_y)_y-i(K_x+K_y)_z,\nn\\
f_z+if_y &=& \phantom{-}(K_x+K_y)_x+i(K_x+K_y)_z,\nn\\
\hat f_z+i\hat f_x &=& -(K_x-K_y)_y+i(K_x-K_y)_z,\nn\\
\hat f_z-i\hat f_y &=& -(K_x-K_y)_x-i(K_x-K_y)_z.
\eer{derivatives_f}

The dual transformation matrices $\tilde U^{(\pm)}$ can now be calculated by the Legendre transformation 
\beq
\tilde U = (K) (U) (K^{-1})
\eeq{dualUagain} 
derived in the previous subsection. Using the partial differential equations for $f$ and $\hat f$ and rewriting the expressions in the semichiral coordinates $\tilde x^i$, the resulting matrices take precisely the form in \re{transformation_matrices_semis}. From the equivalent formulation in \re{transformations_on_semichirals}, one sees that the transformations satisfy $\bbDB{\pm}(\bar\delta^{(\pm)} \tilde x^i )=0$. Hence, the invariance of the dual action  implies
\beq
	\delta \tilde S=0 \quad \Longleftrightarrow \quad \tilde K_{i[j}\tilde U^{(\pm)}{}^{i}_{k]}=0.
\eeq{implies_PDEs}
These partial differential equations are equivalent to the relations in \re{derivatives_f}, and thus we find that the dual action is invariant under the supersymmetry transformations.

To summarize, the original chiral and twisted chiral action is invariant under the $\nff$ supersymmetry if and only if the potential satisfies the Laplace equation. The Laplace equation is dual to equation \re{laplace_dual}, which is equivalent to the Monge-Amp\`ere equation. But this equation is not a sufficient condition for the dual semichiral action to be invariant under the extra supersymmetry. Instead, the dual action is invariant if the transformation parameters in $\Upm$ satisfy certain constraints, equivalent to partial differential equations for the transformation functions in $\tilde U^{(\pm)}$.

%_________________________________________________________________________________________
\subsection{Algebra closure in the dual model}
In \cite{Goteman:2012qk}, it was shown that the $\nff$ transformations on the semichiral fields close to a supersymmetry on-shell provided that the transformation functions satisfy certain partial differential equations. The calculations were straight-forward but tedious. Here, we will see that the on-shell algebra follows directly from the T-duality.

The transformations of the dual coordinates $\tilde x^i$ are derived in \re{dual_susytransformations}-\re{dual_transformation_matrices} from the transformations of the original coordinates $x^\mu$. When discussing the dual action, the original coordinates $x^\mu$ are unconstrained. But since the field equations for the semichiral model are equivalent to the Bianchi identities for the chiral and twisted chiral model, going on-shell by the field equations \re{eqom_dual} is the same as constraining the coordinates $x^\mu$ by the Bianchi identites \re{bianchi_original_(2,2)}. The on-shell closure on the dual coordinates thus follows from the fact that the supersymmetry transformations close on the original coordinates,
\ber
	[\delta(\ep{}_1), \delta(\ep{}_2)]\tilde{x}^i &=& \delta^{i\mu} K_{\mu\nu} [\delta(\ep{}_1), \delta(\ep{}_2)]x^\nu + K_{\mu\nu\tau}\delta_{[1}x^\tau \delta_{2]}x^{\nu}\nn\\
	&=& \delta^{i\mu}K_{\mu\nu}\epb{}_{[2}\ep{}_{1]}i\partial x^\nu\nn\\
	&=& \epb{}_{[2}\ep{}_{1]}i\partial \tilde{x}^i.
\eer{susy_closure_on_dual}
As a summary, the linear supersymmetry transformations on the chiral and twisted chiral fields close off-shell when using the Bianchi identities \re{bianchi_original_(2,2)}, as seen in equation \re{susy_close_on_original}, whereas the non-linear transformations on the semichiral fields close on-shell as seen when using the equivalent field equations \re{eqom_dual}.

%_________________________________________________________________________________________
%_________________________________________________________________________________________
\section{Reduction to $(1,1)$ superspace}
\label{section:reduction}
The original chiral and twisted chiral model reduced to $(1,1)$ superspace is
\beq
	S=\int d^2\xi d^4\theta K(x^\mu) \quad \mathop{\longrightarrow}_{(1,1)}
	\quad S=-\frac{1}{4} \int d^2 \xi d^2\theta D_+ X^a E_{ab} D_- X^b,
\eeq{reduced_chiraltwisted}
where, using the same notation for the $N=(1,1)$ and $\nZZ$ superfields, $\phi\bigr|=\phi$, the coordinates are $X^a=(\phi^c, \chi^t)=(\phi,\bar\phi, \chi, \bar\chi)$. The sum of the metric and the $b$-field,  $E=g+b$, takes the standard form \cite{Lindstrom:2005zr}
\beq
	\begin{array}{ll}
	E_{cc} = K_{cc}+JK_{cc}J, &	E_{ct} = -K_{ct}-JK_{ct}J,\\
		E_{tc} = K_{tc}+JK_{tc}J, &	E_{tt} = -K_{tt}-JK_{tt}J,
	\end{array}
\eeq{Efield_chiraltwisted}
but where the $4\times 4$ matrices in the metric and $b$-field are composed of the $3\times 3$ matrices $K_{\mu\nu}$ as
\beq
	\begin{array}{lcl}
	K_{cc} &=& A_c{}^\mu K_{\mu\nu} A^\nu{}_c,\\
	K_{ct} &=& A_c{}^\mu K_{\mu\nu} A^\nu{}_t,\\
	K_{tc} &=& A_t{}^\mu K_{\mu\nu} A^\nu{}_c,\\
	K_{tt} &=& A_t{}^\mu K_{\mu\nu} A^\nu{}_t,
	\end{array} \quad
	A^\mu{}_c = \left(\begin{array}{cc}1 & 1 \\ 0 & 0 \\ i & -i\end{array}\right), \quad
		A^\mu{}_t = \left(\begin{array}{cc} 0 & 0 \\1 & 1 \\ i & -i\end{array}\right).
\eeq{K_cc_composed}
Hence, the metric and the $b$-field are independent of the coordinate $w\bigr|=i(\phi-\bar\phi-\chi+\bar\chi)$.

Similarly, the semichiral dual model reduced to $(1,1)$ superspace is, after eliminating the auxiliary fields,
\beq
	S=\int d^2\xi d^4\theta K(\tilde{x}^i) \quad \mathop{\longrightarrow}_{(1,1)}
	\quad S=-\frac{1}{4} \int d^2 \xi d^2\theta D_+ \tilde X^a \tilde E_{ab} D_- \tilde X^b,
\eeq{reduced_semichiral}
where the coordinates are $\tilde X^a=(X^L, X^R)=(X_L, \bar X_L, X_R, \bar X_R)$. The sum of the metric and the $b$-field takes the standard form \cite{Lindstrom:2005zr}
\beq
	\begin{array}{ll}
	\tilde E_{LR} = JK_{LR}J + C_{LL}K^{LR}C_{RR},&
	\tilde E_{RL} = -K_{RL}J K^{LR}JK_{RL},\\
	\tilde E_{LL} = C_{LL}K^{LR}J K_{RL},&
	\tilde E_{RR} = -K_{RL}K^{LR}C_{RR},
	\end{array}
\eeq{Efield_semichiral}
where the $4\times 4$ matrices in the metric and $b$-field are composed of the $3\times 3$ matrices $\tilde K_{ij}$ as
\beq
	\begin{array}{lcl}
	K_{LL} &=& \tilde A_L{}^i  \tilde K_{ij} \tilde A^j{}_L,\\
	K_{LR} &=& \tilde A_L{}^i  \tilde K_{ij} \tilde A^j{}_R,\\
	K_{RL} &=& \tilde A_R{}^i  \tilde K_{ij} \tilde A^j{}_R,\\
	K_{RR} &=& \tilde A_R{}^i  \tilde K_{ij} \tilde A^j{}_R,
	\end{array} \quad
	\tilde A^i{}_L = \left(\begin{array}{cc} i & -i \\ i & -i \\ 1 & 1\end{array}\right), \quad
	\tilde A^i{}_R = \left(\begin{array}{cc} i & -i \\-i & i \\ 1 & 1\end{array}\right).
\eeq{K_LR_composed}
Again, the coordinate functions of the metric and the $b$-field are independent of the coordinate parametrizing the direction of the isometry, $\tilde w\bigr|=\tfrac{1}{2}(X_L+\bar X_L-X_R-\bar X_R)$,
\beq
g \sim g_{LL}(\tilde{x},\tilde{y},\tilde{z}) dX_L dX_L + g_{L\bar L}(\tilde{x},\tilde{y},\tilde{z})dX_L d\bar X_L +\dots,
\eeq{metric_nondegenerate}
which does not alter the fact that the metric is non-degenerate.

%_________________________________________________________________________________________
\section{Examples}
\label{section:examples}
\subsection{Flat space}
\label{subsection:Flat space}
To illustrate the results in section \ref{susychapter}, we consider the special case of a quadratic generalized K\"ahler potential. The potential for flat space is
\beq 
K(\phi, \chi)=\frac{1}{2}(\phi +\bar \phi)^2 -\frac{1}{2} (\chi +\bar \chi)^2-\frac{r}{2}\left( (\phi-\bar \chi)^2+(\bar \phi-\chi)^2\right)
\eeq{a}
where $r$ is some arbitrary real constant. 
Gauging the translational isometry, we have
\beq
K=\frac{1}{2} ( x^2- y^2) +\frac{r}{4} \Bigr( z^2 -(x-y)^2 \Bigr) 
\eeq{b}
One can check that this potential satisfies the Laplace equation (\ref{laplacewithisometry}) for any value of $r$. 

Now we perform the duality to semichiral fields to get
\beq
\tilde K=-\frac{1}{2} (\tilde{x}^2 -\tilde{y}^2)-\frac{r}{4} \left(  (\tilde{x}+\tilde{y})^2 +\frac{4 \tilde{z}^2}{r^2}\right).
\eeq{flat potential semis}
In terms of the semichiral coordinates $\bbX{L}$ and $\bbX{R}$, this reads
\ber
\tilde K&=&-\frac{1}{2r}  |\bbX{R}|^2 -\frac{1}{2 r} (1+r^2)|\bbX{L}|^2 \nn\\
&&-\frac{1-r}{2r}(\bbX{L} \bbX{R} +\bbXB{L} \bbXB{R}) -\frac{1+r}{2r} (\bbXB{L} \bbX{R} +\bbXB{R} \bbX{L}).
\eer{flat potential semis X_L}
For any value of $r$, this potential satisfies \re{MongeAmpere} with $c=0$, i.e., the K\"ahler potential will satisfy the Monge-Amp\`ere equation. Therefore, there is no $b$-field. This was to be expected since we have dualized along a translational isometry by equal amounts on $\phi$ and $\chi$. 

We now turn to the supersymmetry transformations and follow the procedure developed in section \ref{susychapter}. Make an ansatz for functions satisfying the partial differential equations in \re{derivatives_f}, 
\ber
	f &=& s x_L^2+i(x+y),\nn\\
	\hat f &=& t x_R^2+ir(x+y)+i(x-y),
\eer{quadraticansatz}
where $s$ and $t$ are two arbitrary constants. The terms multiplying the integration constants $s$ and $t$ in $f$ and $\hat f$ will vanish on-shell. This holds in general; a term $s\cdot g(X_L)$ in $f$ will transform the fields as
\beq
	\delta \tilde x = \epb{+}\bbDB{+}(sg(x_L)) = \epb{+} s g'(x_L)\bbDB{+}x_L = -\epb{+}\frac{s}{2}g'(x_L)\bbDB{+}(\tilde K_{\tilde z}+i\tilde K_{\tilde x}+i\tilde K_{\tilde y}) \mathop{=}_{\tiny\text{on-shell}} 0
\eeq{s_vanishes}
and the same is valid for a term $t\cdot h(x_R)$ in $\hat f$.
The parameters in $\Upm$ are defined in \re{definingalphabeta} and take the constant expressions
\beq
	\begin{array}{lcllcl}
	\alpha &=& \tfrac{1}{2}(-2r+rs+s), \quad & \beta=\tfrac{1}{2}(2r(r+1)+t),\\
	\gamma &=& \tfrac{1}{2}(-2r+rs-s), \quad & \delta=\tfrac{1}{2}(2r(r-1)+t),\\
	\varepsilon &=& -\tfrac{s}{r}, \quad & \kappa=-\tfrac{t}{r}.
	\end{array}
\eeq{alphabetaquadratic} 
The dual transformation matrices can then be derived by Legendre transform and take the form 
\beq
\tilde U^{(+)} = \frac{1}{4}\left(\begin{array}{ccc}
2r-s-rs & -4+2r-rs+s & 2i(-2+\tfrac{s}{r})\\
4+2r-rs-s & 2r-rs+s & 2i(2+\tfrac{s}{r})\\
i(-2r+rs+s) & i(-2r+rs-s) & 2\tfrac{s}{r}
\end{array}\right),
\eeq{dual_transf_matrices_freecase_plus}
and
\beq
\tilde U^{(-)} = \frac{1}{4}\left(\begin{array}{ccc}
-2r(r+1)-t & -4+2r(1-r)-t & 2i(2+\tfrac{t}{r})\\
4+2r(r+1)+t & 2r(r-1)+t & 2i(2-\tfrac{t}{r})\\
i(2r(r+1)+t) & i(2r(r-1)+t) & 2\tfrac{t}{r}
\end{array}\right).
\eeq{dual_transf_matrices_freecase_minus}
The semichiral action \re{flat potential semis} is invariant under these transformations. 
The field equations \re{eqom_dual} in flat space defined by the generalized K\"ahler potential in \re{flat potential semis} take the form
\ber
	(r+1)\bbDB{+}\tilde{x} + (r-1)\bbDB{+}\tilde{y}-\tfrac{2i}{r}\bbDB{+}\tilde{z} &=& 0,\nn\\
	\bbDB{-}\tilde{x}+ \bbDB{-}\tilde{y} -\tfrac{2i}{r}\bbDB{-}\tilde{z} &=&0.
\eer{eqom_freecase}
The integration constants $s$ and $t$ in the transformations multiply field equations and vanish when \re{eqom_freecase} are used. Using the field equations, one can then check explicitly that the transformation defined by the matrices in \re{dual_transf_matrices_freecase_plus}-\re{dual_transf_matrices_freecase_minus} close to a supersymmetry on-shell.

%------------------------------------------------------------------------------
\subsection{Non-quadratic potential}
Non-flat generalized K\"ahler potentials can also be constructed. One example is\footnote{The author thanks Ulf Lindstr\"om for pointing out related examples in \cite{Bogaerts:1999jc}.}
\beq
	K(x, y, z) = z\cdot\bigr(F(x+iy)+\bar F(x-iy)\bigr)
\eeq{nonflat_K}
The potential satisfies the Laplace equation \re{laplacewithisometry}, hence the original chiral and twisted chiral sigma model has $\nff$ supersymmetry off-shell.
As the functions $F$, $\bar F$ one can consider, for example, $F=(x+iy)^2$, so that the original Lagrangian takes the qubic form 
\beq
	K(x, y, z)=	z\cdot \bigr(x^2-y^2\bigr).
\eeq{nonflat_K_example}
The Legendre transform \re{varyingpotentials}-\re{dualmodel} gives
\beq
\begin{array}{lclcl}
	K_{x} & =& \phantom{-}2z\cdot x & =& \tilde{x},\\
	K_{y} & = & -2z\cdot y & =& \tilde{y},\\
	K_{z} & = & x^2-y^2 & =& \tilde{z},
\end{array}
\eeq{legendre_example}
where $x^\mu=(x,y,z)$ are unconstrained. This implies that the dual generalized K\"ahler potential takes the form
\beq
	\tilde K(\tilde{x},\tilde{y},\tilde{z}) = - \sqrt{\tilde{z}(\tilde{x}^2-\tilde{y}^2)}.
\eeq{nonflat_dualK}
This potential satisfies \re{laplace_dual}, which is equivalent to the Monge-Amp\`ere equation, hence the dual describes hyperk\"ahler geometry. The dual potential in semichiral coordinates reads
\beq
	\tilde K(\bbX{L}, \bbXB{L}, \bbX{R}, \bbXB{R}) = -\sqrt{\tfrac{1}{2} (\bbX{L}+\bbXB{L}+\bbX{R}+\bbXB{R})i(\bbX{L}-\bbXB{L})i(\bbX{R}-\bbXB{R})}.
\eeq{nonflat_dualK_semichiral}
The determinant of the Hessian corresponding to this Lagrangian is $\hbox{det}\, \tilde K= 1/(8\tilde{z})$.

We now turn to the supersymmetry transformations. Making an ansatz for $f$ and $\hat f$ to be quadratic, in order to satisfy the partial differential equations in \re{derivatives_f} they must be of the form
\ber
f &=& 2\bigr(sx_L^2+2xy-i(x+y)z\bigr)\nn\\
\hat f &=& 2\bigr(tx_R^2+2xy+i(x-y)z\bigr),
\eer{ansatz_f}
where $s$ and $t$ are some arbitrary integration constants. Again, the terms multiplying $s$ and $t$ in the transformations $\delta \tilde x^i$ will vanish on-shell. With these functions, the transformation parameters take the form in \re{definingalphabeta}. For clarity, we display here only the on-shell part of the transformations,
\beq
\begin{array}{lcllcl}
\alpha &=& 2iy\frac{x^2+y^2}{(x^2-y^2)z},\quad&
	\beta &=& 2iy\frac{x^2+y^2}{(x^2-y^2)z},\\
\gamma &=& 2ix\frac{x^2+y^2}{(x^2-y^2)z},&
	\delta &=& 2ix\frac{x^2+y^2}{(x^2-y^2)z},\\
	\varepsilon &=&-\frac{4xy}{x^2-y^2},\quad &
	\kappa &=& -\frac{4xy}{x^2-y^2}.
\end{array}
\eeq{identify_alpha_example}
The supersymmetry transformations for the semichiral model can now be derived by the Legendre transform $\tilde U = (K)(U)(K^{-1})$ and take the form
\beq
	\tilde U^{(+)} = \left(\begin{array}{ccc}
	\frac{2i\tilde y(\tilde x^2+\tilde y^2)\tilde z}{(\tilde x^2-\tilde y^2)^2} & -1-\frac{2i\tilde x(\tilde x^2+\tilde y^2)\tilde z}{(\tilde x^2-\tilde y^2)^2} & -i-\frac{2i\tilde x\tilde y}{\tilde x^2-\tilde y^2} \\
	1+\frac{2i\tilde y(\tilde x^2+\tilde y^2)\tilde z}{(\tilde x^2-\tilde y^2)^2} & -\frac{2i\tilde x(\tilde x^2+\tilde y^2)\tilde z}{(\tilde x^2-\tilde y^2)^2} &  -i-\frac{2i\tilde x\tilde y}{\tilde x^2-\tilde y^2}\\
	\frac{2\tilde y(\tilde x^2+\tilde y^2)\tilde z}{(\tilde x^2-\tilde y^2)^2} & -\frac{2\tilde y(\tilde x^2+\tilde y^2)\tilde z}{(\tilde x^2-\tilde y^2)^2} & -i-\frac{2i\tilde x\tilde y}{\tilde x^2-\tilde y^2}
	\end{array}\right)
\eeq{Utilde_nonquadratic}
and similar for $\tilde U^{(-)}$.
One can explicitly check that, for arbitrary values of the integration constants $s$ and $t$ (not displayed in \re{Utilde_nonquadratic} since this is the on-shell part only), 
 $\bbDB{\pm}(\bar\delta^{(\pm)} \tilde x^i )=0$ and that the partial differential equations $\tilde K_{i[j}\tilde U^{(\pm)}{}^{i}_{k]}=0$ are satisfied, hence the semichiral action with generalized K\"ahler potential \re{nonflat_dualK} is invariant under these supersymmetry transformations.

%_________________________________________________________________________________________
\section{Summary and conclusions}
In this paper, the T-duality between four-dimensional chiral and twisted chiral models and semichiral models has been investigated. 
Whereas the chiral and twisted chiral model admints off-shell $\nff$ supersymmetry if and only if the generalized K\"ahler potential satisfies the Laplace equation \cite{Gates:1984nk}, the semichiral model can only admit on-shell $\nff$ supersymmetry \cite{Goteman:2009ye}, \cite{Goteman:2012qk}. 

What happens when one starts with a chiral and twisted chiral model with off-shell $\nff$ supersymmetry and dualize into a semichiral model? Will the transformations be satisfied on-shell, and do we find additional constraints on the generalized K\"ahler potential? How can the non-linear on-shell $\nff$ transformations of the semichiral fields be related to the linear off-shell transformations of the chiral and twisted chiral fields? These were the main questions we wanted to address in this work.

We find that, in order to dualize into a semichiral model with $\nff$ supersymmetry, additional non-linear terms must be added to the chiral and twisted chiral supersymmetry transformations. These terms are of the kind that they vanish when the chirality constraints are used, and so do not appear for the chiral and twisted chiral model. In other words, they vanish when the Bianchi identities are used in the original chiral and twisted chiral model, or, equivalently, when the field equations are used in the dual semichiral model,
\beq
	\bar{\delta}^{+} x = \epb{+}[\bbDB{+} y +i\alpha \bbDB{+}x_L] \quad\mathop{\longrightarrow}_{\text{Bianchi}} \quad \epb{+}\bbDB{+}y.
\eeq{summarybianchi}
The additional non-linear terms hence vanish when considering the on-shell supersymmetry algebra for the semichiral fields, and the algebra closes on-shell as a direct consequence of the T-duality. But the terms are crucial for the invariance of the semichiral action under the transformations.

The supersymmetry transformation matrices $\tilde U$ for combinations of the the semichiral fields can be calculated from the chiral and twisted chiral transformations $U$ by Legendre transform,
\beq
	\tilde U = (K) (U) (K^{-1}).
\eeq{summaryUtilde}
Even though the underlying system is a four-dimensional target space parametrized by semichiral fields $(\bbX{L}, \bbXB{L}, \bbX{R},\bbXB{R})$, the T-duality only provides the supersymmetry transformations for the three combinations $(\tilde x, \tilde y, \tilde z)$ of the semichiral fields.
We define the parameters in $U$ such that the resulting transformations agree with the transformations on the semichiral fields obtained recently in \cite{Goteman:2012qk}, and we check that the transformations close to a supersymmetry on-shell and leave the action invariant. 

%_________________________________________________________________________________________
\section*{Acknowledgements}
The author is indebted to Marcos Crichigno, Ulf Lindstr\"om and Martin Ro\v{c}ek for invaluable support, comments and inspiration. 

Parts of this research project was carried out when the author was at Stony Brook, NY. The author would like to thank Stony Brook University for hospitality, and the equality grant from the Dept.~of Physics \& Astronomy, Uppsala University, for funding the stay.

%------------------------------------------------------------------------------
\appendix

\section{The project in a nutshell}
\label{appendixA}
Summarizing the most relevant equations and comparing the two dual models.
\begin{center}
\begin{tabular}{p{8cm} p{8cm}}
\bf{Original model:} & \bf{Dual model:}\\[3mm]
$S=\int K(x^\mu)=\int K(x, y, z)$ &  $\tilde S=\int \tilde K(\tilde{x}^i)=\int K(\tilde{x},\tilde{y},\tilde{z})$\\[3mm]
where 
$\left\{\begin{array}{lcl}x &=& \phi+\bar\phi \\
y &=& \chi+\bar\chi \\
z &=&i(\phi-\bar\phi+\chi-\bar\chi) 
\end{array}\right.$ 
&
where 
$\left\{\begin{array}{lcl}\tilde{x} &=& \tfrac{i}{2}(\bbX{L}-\bbXB{L}+\bbX{R}-\bbXB{R}) \\
\tilde{y} &=& \tfrac{i}{2}(\bbX{L}-\bbXB{L}-\bbX{R}+\bbXB{R}) \\
\tilde z &=&\tfrac{1}{2}(\bbX{L}+\bbXB{L}+\bbX{R}+\bbXB{R}) 
\end{array}\right.$\\[3mm]
$K_\mu = \delta_{\mu i}\tilde{x}^i$ & $\tilde K_i = -\delta_{i\mu}x^\mu$\\
$K_{\mu\nu} = -\delta_{\mu i}(\tilde K^{-1})^{ij}\delta_{j\nu}$ & $\tilde K_{ij} = -\delta_{i\mu}(K^{-1})^{\mu\nu}\delta_{\nu j}$\\
$\bbD{}x^\mu = -\delta^{\mu i}\tilde K_{ij}\bbD{}\tilde{x}^j$ &
$\bbD{}\tilde{x}^i = \delta^{i \mu} K_{\mu\nu}\bbD{}x^\nu$\\[3mm]
\end{tabular}

\begin{tabular}{p{8cm} p{8cm}}
Bianchi identities: & Bianchi identities: \\
$\bbDB{+}(z+ix + iy)=0$ & $\bbDB{+}\bbDB{-}(\tilde{z}-i\tilde{x})=0$\\
$\bbDB{-}(z+ix - iy)=0$ & $\bbDB{+}\bbD{-}(\tilde{z}-i\tilde{y})=0$\\[3mm]
Field equations: & Field equations: \\
$\bbDB{+}\bbDB{-}(K_{z} - i K_x)=0$ & $\bbDB{+}(\tilde K_{\tilde{z}}+i\tilde K_{\tilde{x}} +i\tilde K_{\tilde{y}})=0$\\
$\bbDB{+}\bbD{-}(K_{z} - i K_y)=0$ & $\bbDB{-}(\tilde K_{\tilde{z}}+i\tilde K_{\tilde{x}} -i\tilde K_{\tilde{y}})=0$\\[3mm]
\end{tabular}

\begin{tabular}{p{8cm} p{8cm}}
Supersymmetry: & Supersymmetry:\\
$\delta x^\mu = \epb{\alpha}U^{(\alpha)}{}^\mu_\nu \bbDB{\alpha}x^\nu + c.c.$ &
$\delta \tilde{x}^i = \epb{\alpha}\tilde U^{(\alpha)}{}^i_j \bbDB{\alpha}\tilde{x}^j + c.c.$\\
$U^{(\alpha)}$ constant $3\times 3$ matrices &
$\tilde U = (\partial \partial K)U (\partial\partial K)^{-1}$ not constant\\
susy algebra closes off-shell & susy algebra closes on-shell\\
(using Bianchi identities) & (using field equations)\\[3mm]
\end{tabular}

\begin{tabular}{p{8cm} p{8cm}}
Fourth coordinate: & Fourth coordinate:\\
$w=i(\phi-\bar\phi-\chi+\bar\chi)$ & $\tilde w= \tfrac{1}{2}(\bbX{L}+\bbXB{L}-\bbX{R}-\bbXB{R})$\\[3mm]
\end{tabular}

\begin{tabular}{p{8cm} p{8cm}}
Invariance of action: & Invariance of action:\\
	$	K_{\mu[\nu}U^{(\alpha)}{}^\mu_{\rho]}\bbDB{\alpha}x^\nu \bbDB{\alpha}x^\rho =0$ & $\tilde K_{i[j}\tilde U^{(\alpha)}{}^i_{k]} \bbDB{\alpha}\tilde{x}^j \bbDB{\alpha}\tilde{x}^k =0$\\
	$\Leftrightarrow 	K_{x x}+K_{y y} + 2K_{zz} = 0$ &
	$\Leftrightarrow$ PDEs for $f, \hat f$\\[3mm]
\end{tabular}

\begin{tabular}{p{8cm} p{8cm}}
Reduced to $(1,1)$: & Reduced to $(1,1)$:\\
$S=\int d^2\xi d^2\theta D_+ X^a E_{ab} D_- X^b$ &
$\tilde S=\int d^2\xi d^2\theta D_+ \tilde X^a \tilde E_{ab} D_- \tilde X^b$\\
$X^a=(\phi^c, \chi^t) $ & $\tilde X^a=(X^L, X^R)$\\
$K_{ab} = A_a{}^\mu K_{\mu\nu} A^\nu{}_b$ &
$\tilde K_{ab} = \tilde A_a{}^i \tilde K_{ij} \tilde A^j{}_b$.
\end{tabular}
\end{center}

\end{document}